\documentclass[aps,pre,twocolumn,groupedaddress,showpacs,floatfix]{revtex4}
\usepackage{amsmath}
\usepackage{amsfonts}
\usepackage{amssymb}
\usepackage{graphicx}
\usepackage{subfigure}
\begin{document}
\title{Pulse Propagation in Chains with Nonlinear Interactions}
\author{Alexandre Rosas and Katja Lindenberg}
\affiliation{
Department of Chemistry and Biochemistry,
and Institute for Nonlinear Science
University of California San Diego
La Jolla, CA 92093-0340}

\begin{abstract}
Pulse propagation in nonlinear arrays continues to be of interest
because it provides a possible mechanism for energy
transfer with little dispersion.  Here we show that common measures of
pulse dispersion might be misleading; in strongly anharmonic systems
they tend to reflect a succession of extremely narrow pulses traveling at
decreasing velocities rather than the actual width of a single pulse.
We present analytic estimates for the fraction of the initial energy
that travels in the leading pulses.  We also provide analytic
predictions for the leading pulse velocity in a Fermi-Pasta-Ulam
$\beta$-chain.

\end{abstract}

\maketitle

The stability of localized energy, e.g. in the form of breathers,
in translationally invariant
nonlinear arrays, and the way in which localized energy packets can be
transported in these arrays, has been a topic of interest for several
decades, and continues to be of great interest for a number of reasons.
One is that many of the ideas on the subject have recently and
increasingly been confirmed experimentally.  Another is the possible
importance of the subject in the transport of energy in biological
systems.  A third is the ever increasing numerical capability that
allows simulations of larger systems over longer times.  A recent
focus issue of the journal \emph{Chaos}~\cite{chaos}
contains some of the most current 
contributions and reviews of the subject, and covers the three topics
just mentioned.  

While the advances of the past few years are exciting and
enormously instructive, the analytic understanding of these phenomena
has been made difficult by the fact that the systems are nonlinear.
Many of the available results (including those obtained in our group)
are numerical, and it is sometimes difficult and even misleading
to generalize from these results (for reviews of the subject
preceding the special issue
noted above, see~\cite{flach,aubry} and references therein). 
Our contribution in this
paper is an analytic understanding of results previously obtained only
numerically.

A typical set of questions that one can pose is the following: Suppose
that a single unit in a nonlinear array is given an initial velocity.
How will this velocity/energy propagate through the array?  Will some
or all of the energy remain localized, or will it spread? If a localized
moving pulse does develop, at what velocity will it propagate?  These
are some of the signal propagation issues that we address analytically.

We focus on the one-dimensional Fermi-Pasta-Ulam (FPU)-type
problem for unit mass particles described by the Hamiltonian 
\begin{equation}
H=\sum_i \frac{\dot{x}_i^2}{2} +\sum_i V(x_i-x_{i-1}),
\end{equation}
where $x_i$ is the displacement of particle $i$ from its equilibrium
position, and $V(z)$ is the potential
\begin{equation}
V_n(z) = \frac{k'}{n}\sum_{i}|z|^n +\frac{k}{2}\sum_{i}z^2.
\end{equation} 
For the FPU $\beta$-problem $n=4$, but we retain $n$ as a
general power because a
number of theories and simulations deal with other values of $n$, and
portions of our analysis do as well.  The
parameters $k$ and $k'$ are the harmonic and anharmonic force constants,
respectively.
The variables and the time
can be scaled so that the only distinct cases of this problem
are $k=0$ (purely anharmonic chain), $k'=0$ (purely harmonic chain),
and $k, k' \neq 0$ (``mixed'' chain; $k=k'=1$ is a convenient choice).  
The control parameter is then the initial velocity.
The equation of motion for the $i^{th}$ particle in a mixed chain then is
\begin{equation}
\begin{split}
\ddot{x}_i &=  \left| x_{i+1} - x_i\right|^{n-1}
\mathrm{sgn} (x_{i+1} - x_i) + ( x_{i+1} - x_i) \\
&+  \left| x_i - x_{i-1} \right|^{n-1}
\mathrm{sgn} (x_i - x_{i-1}) + ( x_i - x_{i-1} ),
\label{eq:motion_scaled}
\end{split}
\end{equation}
where $\mathrm{sgn} (x)=\pm 1$ for $ x \gtrless 0 $. 
Initially all particles are at rest
in their equilibrium positions except for one particle (far from any
boundaries) that has initial velocity $v_0$.
We take
the chain to be sufficiently long and the boundaries sufficiently far from
the initial excitation that their precise nature does not matter for our
analysis.

In~\cite{sarmiento}, Sarmiento {\em et al.} analyze the pulse evolution
in terms of the mean distance from the initial site (``pulse
position'') and its dispersion (``pulse width''),
\begin{equation}
\langle x\rangle = \frac{\sum_i \left|i\right| E_i}{\sum_i E_i}, \quad
\sigma^2 = \frac{\sum_i i^2 E_i}{\sum_i E_i} - \langle x
\rangle ^2.
\label{eq:md}
\end{equation} 
The local energy is defined as
\begin{equation}
E_i = \frac{\dot{x}_i^2}{2} + \frac{1}{2} V(x_{i+1}, x_i)
+ \frac{1}{2} V(x_i, x_{i-1}).
\label{eq:localenergy}
\end{equation} 

\begin{figure}
\begin{center}
\includegraphics[width=7cm]{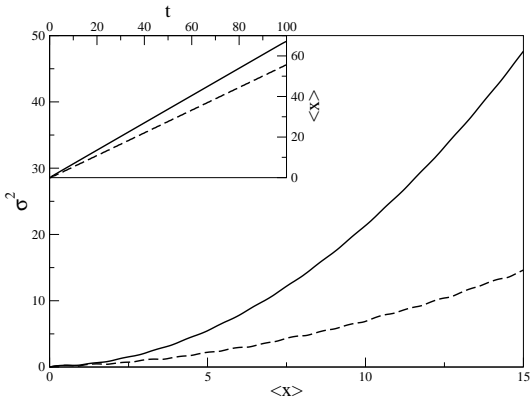}
\end{center} 
\caption{Mean distance and dispersion of the pulse as defined in
Eqs.~\eqref{eq:md} for a purely quartic
potential (dashed line) and a mixed potential with $n=4$ and
$v_0=1.0$ (solid line).
\label{fig:spread}}
\end{figure} 

We begin by considering the spreading of the initial pulse. 
The width of the pulse as time proceeds is often
invoked as a measure of the ability of the nonlinearity to keep the
energy localized.  It is well known that in a harmonic lattice
an initial pulse spreads even as it moves.  
In a mixed chain, one expects less spreading
for a higher initial velocity $v_0$ since a
more energetic pulse samples the more anharmonic portions of the
potential.  Indeed, highly localized breathers have been shown to be
exact solutions for the purely anharmonic chain in the limit
$n\to\infty$~\cite{page}.  For a quartic anharmonicity ($n=4$) the
contribution of the harmonic and anharmonic contributions to the potential
energy are equal at the maximum displacement associated with kinetic
energy $v_0^2/2 = 4$. 
Therefore when $v_0\ll \sqrt{8}$ ($v_0 \gg \sqrt{8}$) the dominant
contribution to the potential energy of the pulse is the harmonic
(anharmonic) portion.

In Fig.~\ref{fig:spread} we show the pulse width as a
function of pulse position for $n=4$ in two cases. In one, 
$v_0=1$ so that the harmonic portion of the potential is strongly sampled
by the excitation. In the other, $v_0\to\infty$, i.e., the potential is
essentially a purely quartic potential. 
The pulse in the purely anharmonic potential
is more localized after traveling a given mean distance than is the pulse
in the mixed system~\cite{sarmiento}.   The inset shows the mean pulse
position as a function of time (the pulses move at a constant speed).

\begin{figure}
\begin{center}
\subfigure[\label{fig:snap-almostharm}]
{\includegraphics[width=8cm]{fig2a.eps}}
\vskip 8pt
\subfigure[\label{fig:snap-quart}]
{\includegraphics[width=8cm]{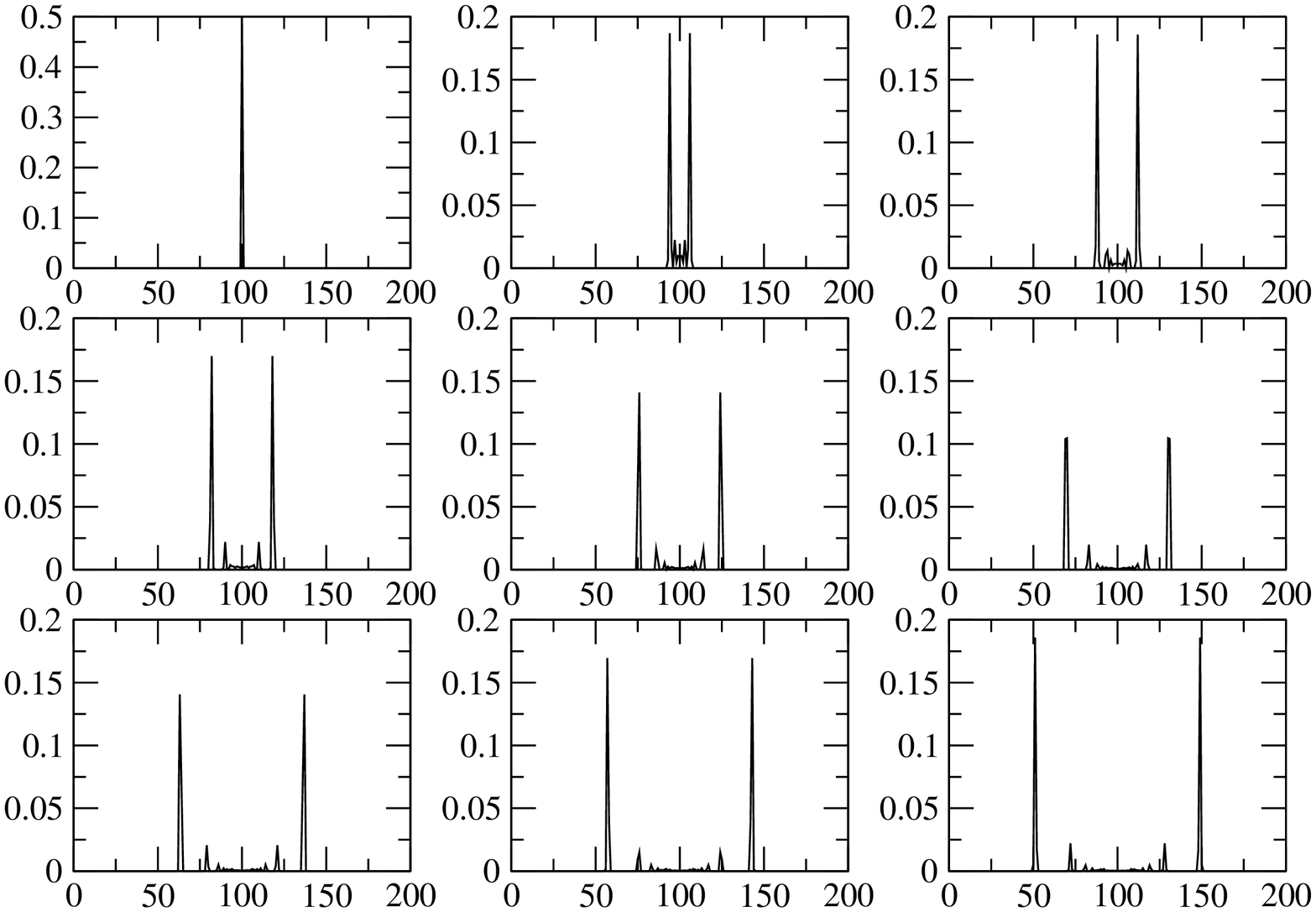}}
\end{center} 
\caption{Snapshots of the normalized local energy profile vs lattice site
for the mixed potential with $n=4$ and initial pulse velocity
$v_0=1$ (a) and $v_0\to\infty$ (b). From left to right and top to bottom, time
runs from zero
to 80 in steps of ten (adimensional) units.}
\end{figure} 

We note that the lower energy pulse 
moves more rapidly than the higher energy pulse.
This result may appear contradictory with those obtained
earlier in~\cite{sarmiento} where, \emph{for a given initial pulse energy},
pulse velocities in purely harmonic and purely anharmonic systems
were compared.  A pulse in a purely
harmonic system moves at a speed that is independent of the initial
velocity $v_0$, while the pulse speed in a purely quartic system
increases with increasing $v_0$~\cite{sarmiento}.   Therefore, for
sufficiently high initial pulse velocity, 
a pulse in a purely harmonic system moves {\em more slowly}
than in a purely anharmonic chain~\cite{sarmiento}. 
In a mixed chain, on the other hand, the pulse speed as a function of
$v_0$ is bounded
\emph{below} by the higher of the two (purely harmonic and purely
quartic), approaching the purely harmonic behavior at low $v_0$ and
the purely quartic behavior at large $v_0$.

Alternatively but equivalently, one can say
that the velocity of a pulse in a purely harmonic chain is independent of
its amplitude while in an anharmonic chain the pulse speed decreases with
decreasing amplitude. This observation has important implications in the
understanding of the way in which the pulse width increases as the pulse
travels away from the site of origin.  To see this, consider a
low-velocity pulse launched in a mixed chain.  The harmonic portion
of the potential here dominates the evolution of the pulse,
as shown in
Fig.~\ref{fig:snap-almostharm}.  Two symmetric fronts travel away from
the origin carrying part of the energy and spreading.  The remaining
energy is progressively distributed among the particles between the
pulses. Altogether one observes a gradual broadening of the two
pulses, exactly as one imagines a pulse broadening to occur.

However, for a mixed potential with high initial pulse velocity
the situation is quite different, as can be
seen in Fig.~\ref{fig:snap-quart}. Now a portion of the energy
travels symmetrically away from the center in extremely localized
pulses that in fact \emph{remain highly localized}.  The remaining energy
is successively ``launched'' from the origin in the form of secondary
pulses of smaller and smaller amplitude which travel more and more
slowly (the low initial velocity or nearly harmonic case can be thought of
in these terms as well, but
the secondary pulses travel at essentially the same speed as the
primary pulses).
Hence, there is a series of narrow pulses of decreasing amplitude that are
getting further apart from one another, giving rise to the apparent
``dispersion''.  Moreover, the maxima of the pulses oscillate;
when the energy is more concentrated in one particle
the maximum is higher than when it is shared between two or more.

The extreme localization of the pulses suggests that a two- or
three-particle approximation may capture the essence of the physics of
the problem, and it is this feature that we use to arrive at analytic
results.  The point to stress here is that \emph{the second moment}
$\sigma^2$
\emph{is in fact a deceptive measure of the dispersion of the pulse}
except in an essentially harmonic system.  

We begin by estimating the energy in the primary pulses and, from this,
the pulse velocity as determined by the pulse energy.
We assume that the only effects of the restoring forces are to split
part of the energy into two pulses, and to keep
the remainder of the energy at the origin, from where 
it will create the secondary pulses.
Therefore we need to calculate how much energy is transmitted
from the particle at $i=0$ to the particles at $i=\pm 1$, and how quickly
it is transmitted.

The first step is achieved by considering a three particle
system and neglecting the rest of the chain. This approximation
presupposes that the potential is sufficiently steep that the
particles at $i=\pm 2 $ barely move before the particles at $i=\pm 1$
have acquired their full velocity. Obviously, this is not strictly
true for finite $n$; however, we will show that it is a very good
approximation, especially for high pulse velocities.

\begin{figure}
\begin{center}
\includegraphics[width=7cm]{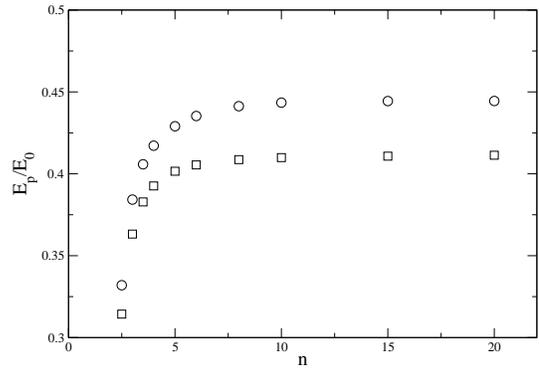}
\end{center} 
\caption{Relative energy in the primary pulse as a function of
the nonlinearity
$n$ in the potential.  Circles: purely anharmonic potential.
Squares: mixed potential with $v_0=10$. 
\label{fig:pulseenergy}}
\end{figure} 

Initially, the three particles are in their equilibrium positions
and all of the energy $E_0=v_0^2/2$ is concentrated in the middle particle,
$i=0$.
Some of the energy is transfered to the neighbors as the springs
compress and stretch, and at some time later
the energy of the three particles is once again all kinetic.  The
symmetry of the system requires that this occur when the
particles at $i=\pm 1$ acquire their maximum velocity, $u$
(equal by symmetry), and the one at
$i=0$ its minimum velocity, $u_0$. 
The three-particle system
oscillates back and forth between these two configurations, but we are
only interested in this first portion of the cycle.
Energy and momentum conservation lead to 
\begin{equation}
u=\frac{2}{3} v_0, \qquad u_0 = -\frac{1}{3} v_0.
\label{eq:prediction}
\end{equation} 
Each primary pulse therefore carries away an energy
\begin{equation}
E_p=\frac{1}{2}u^2 = \frac{2}{9} v_0^2 = \frac{4}{9}E_0.
\end{equation}

Figure~\ref{fig:pulseenergy} shows simulation results for the primary pulse
energy as a function of the power $n$ of the potential for two cases. 
The circles correspond to an initial pulse velocity $v_0\to\infty$ (or,
alternatively, a purely quartic potential with any $v_0$).  The squares
correspond to an initial pulse velocity $v_0=10$.  Both lie in the
regime where the dynamics is dominated by the anharmonicity, $v_0\gg
\sqrt{8}$, but our theory is expected to improve with increasing $v_0$.
The asymptotic value for the high $v_0$ case is
$E_p/E_0=4/9$, exactly as predicted.  The agreement is very good
even for $n=4$, where our prediction is already within a few percent of
the correct value.  For the lower initial pulse velocity the energy ratio
is asymptotically only about 7\% smaller than predicted, a reflection of the
fact that as the harmonic component becomes more important
the pulse occupies more than three sites.  

Next we turn our attention to the calculation of the velocity of the
primary pulse when $n=4$ (the FPU case). For this calculation we
simplify our model
even further 
and consider only a two-particles system, one of which has the initial
velocity $u=2v_0/3$. We then calculate the
time, $T(v_0)$, for the second particle to acquire the same velocity
as the first, i.e., the time at which the velocities of the two
particles are equal. We maintain that this is the time necessary for the
primary pulse to travel from one particle to the next.  To calculate this time
it is not necessary to actually integrate the equations of motion.
Defining $z = x_1 - x_2$, we have for $n=4$,
$\ddot{z} = -2 z^3 - 2z$.
This equation of motion is derived from a potential $V(z) = z^4/2
+ z^2$,
and the initial conditions are $z(0) = 0$ and $ \dot{z}(0) = u$.
From energy conservation we have that the final energy
(kinetic plus potential) is equal to the initial energy (all kinetic), 
\begin{equation}
\frac{1}{2} \dot{z}^2 + V(z)
=\frac{1}{2}u^2, 
\label{eq:velpos}
\end{equation} 
from which it follows that
$\dot{z} = \sqrt{ u^2 -2 z^2 - z^4}$.
Furthermore, the two particles have the same velocity when
$ \dot{z}=0$.  It immediately
follows that the relative displacement at that moment is given by
$z_m = \sqrt{\sqrt{1+u^2}-1}$.
Now we can integrate $\dot{z}$,
\begin{equation}
T(v_0) = \frac{1}{z_m} \int_0^1 \frac{\mathrm{d} \eta}
{\sqrt{\left (1 - \eta^2 \right )\left (\coth^2(\phi/2) + \eta^2 \right )}},
\end{equation} 
where we have introduced the variables $\eta \equiv z/z_m$ and $\phi =
\sinh^{-1}(u)$.
This integral can be done analytically~\cite{grad}:
\begin{equation}
T(v_0) = \frac{1}{\sqrt{2}\cosh^{1/2}(\phi)}
K\left (\frac{ \left (\cosh(\phi) - 1\right)^{1/2}}
{\sqrt{2}\cosh^{1/2}(\phi)} \right ),
\end{equation} 
where $ K(x) $ is the complete elliptic integral of the first
kind~\cite{grad}. 
The pulse velocity is just the inverse of this time:
\begin{equation}
C(v_0) = T^{-1}(v_0).
\label{eq:pulsevelocity}
\end{equation} 
For $v_0\gg 1$ (strongly anharmonic potential) one finds from dimensional
analysis~\cite{ourchain} that $C(v_0)\sim v_0^{1/2}$.

\begin{figure}
\begin{center}
\includegraphics[width=7cm]{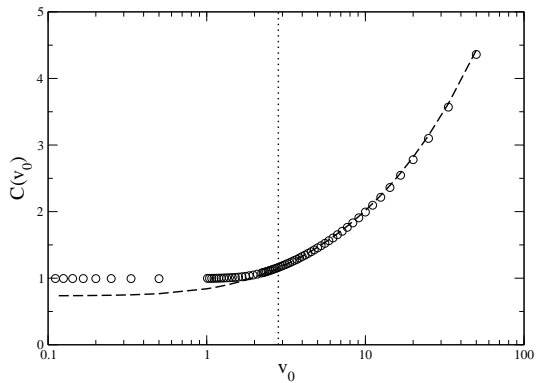}
\end{center} 
\caption{
Pulse velocity vs initial pulse velocity.
The circles are the simulation results for a full chain, and the broken
line is Eq.~(\ref{eq:pulsevelocity}).  
The dotted line corresponds to 
$v_0=\sqrt{8}$.
\label{fig:pulsevel}}
\end{figure}

In Fig.~\ref{fig:pulsevel}, we compare the results of our
approximation with the numerical simulation of the full chain as a function
of $v_0$.  The circles are the simulation results, and the 
broken line is our two-particle approximation,
Eq.~(\ref{eq:pulsevelocity}).
The agreement is clearly excellent for initial pulse velocities above
$v_0 > \sqrt{8}$, the value that we offered as a limit for the validity of
this approach.

In this communication we have shown that traditional measures of pulse
propagation in arrays with nonlinear interactions may
be misleading.  We discussed systems with interactions that have both
harmonic and anharmonic contributions, and argued that the behavior of a
pulse launched by imparting an initial velocity $v_0$ to a particle at
one site in such a system depends strongly on $v_0$.
For low velocities the harmonic portions
of the potential are primarily sampled, and the pulse behaves as it
would in a harmonic system. For high velocities the anharmonic
portions of the potential dominate the behavior, and the pulse
propagates as it would in a purely anharmonic chain. It is in the
anharmonic regime that one must view traditional measures
of pulse propagation with some caution.  
In particular, we have shown
that in the anharmonic regime the usual second moment
``pulse width'' is not a measure of the way a single pulse spreads, but
rather, of the span covered by a series of very narrow pulses of
decreasing velocity. In a statistical measure this appears as a growing
second moment.  We have also presented analytic
estimates for the energy and velocity of the leading pulse, and have
shown by comparison with numerical simulations
that our estimates are extremely accurate in the anharmonic
regime. 

A number of extensions of our approach are possible, albeit with some
analytic complications. For example, the approach can be extended to the
FPU $\alpha$-model that includes cubic as well as quartic interactions.
Some of the polynomial solutions and integrations that we have carried
out analytically might then have to be done numerically.
The same is true were one to include dissipation.
The model can be extended to include a local harmonic potential, and
also to higher dimensions.  We are currently exploring these
generalizations.

\vskip 5pt
%\section*{Acknowledgments}
This work was supported by the Engineering Research Program of
the Office of Basic Energy Sciences at the U. S. Department of Energy
under Grant No. DE-FG03-86ER13606.

\end{document}